\documentstyle[preprint,aps]{revtex}
\input epsfig.sty
\newcommand{\be}{\begin{equation}}
\newcommand{\la}{\label}
\newcommand{\ee}{\end{equation}}
\newcommand{\bea}{\begin{eqnarray}}
\newcommand{\eea}{\end{eqnarray}}
\begin{document}

\title{The Schr\"{o}dinger particle in an oscillating spherical cavity}
\author{K.~Colanero and M.~-C.~Chu}
\address{Department of Physics, The Chinese University of Hong Kong, Shatin,
N.T., Hong Kong.}
\maketitle
\tightenlines

\begin{abstract}
We study a Schr\"{o}dinger particle in an infinite spherical well with an
oscillating wall.  Parametric resonances emerge 
when the oscillation frequency is equal to the energy difference between two
eigenstates of the static cavity.  Whereas an analytic calculation
based on a two-level system approximation reproduces the numerical results
at low driving amplitudes $\epsilon$, we observe a drastic change of
behaviour when $\epsilon > 0.1$, when new resonance states appear bearing
no apparent relation to the eigenstates of the static system.
\end{abstract}

\vspace{1 cm}
\nopagebreak
We study in this article the behaviour of a Schr\"odinger particle
confined in a spherical cavity with an oscillating boundary that 
constitutes 
a particular kind of time-dependent perturbation. 
Our study provides a conceptually simple ``laboratory'' in which the
subtle and nontrivial aspects of the resonant coupling between the
oscillating wall and a particle trapped inside the
cavity can be investigated.  
Our original motivation in this work comes from our attempt to
construct a  dynamical bag model of hadrons \cite{bag}; however,
our results may bear implications on the physics of a wide range 
of systems such as 
cavity QED \cite{em} and perhaps even sonoluminescence \cite{sono}.

The system of a one-dimensional vibrating perfect
cavity with quantized electromagnetic fields has been well studied \cite{em}.
It was found that  the electromagnetic field energy density inside
a cavity vibrating at one of its resonance
frequencies concentrates into narrow peaks regardless of the detailed
trajectories of the oscillating cavity wall \cite{law,cole,ywu}.
Furthermore, the amplitudes of these energy
wave packets grow rapidly in time, producing sharp and intense pulses
of photons. The distortion of the vacuum fields arising from the 
cavity wall motions leads to dynamical modifications of the Casimir
effects \cite{jae}, which represents a fundamentally important and 
interesting feature of quantum physics.
The problem of a quantum particle in a box with moving walls 
has also been studied with an analytical approach \cite{dodonov2}, 
but the possibility of resonances was not discussed, 
which is the main interest in this work.

If the oscillation amplitude $\epsilon R_0$ is small compared to the original
cavity radius $R_0$, perturbation theory can be used to
calculate the transition amplitudes between two states of the
unperturbed system. This corresponds to what is usually observed in
experiments.  However the non-perturbative solutions of
the complete time-dependent Hamiltonian ($H=H_0+H_1(t)$), 
where $H_0$ is the time-independent part of the Hamiltonian,
can in principle be remarkably different from the
perturbative ones and can give rise to non-trivial features.

We consider, as a first step, an infinite spherical well 
with oscillating walls:
\be
V(r)=\left\{ \begin{array}{ll}
0 & \mbox{if $r < R(t)$}\\
\infty & \mbox{if $r \geq R(t)$}
\end{array}
\right. ,
\la{d.pot}
\ee
where $R(t)=R_0 (1 + \epsilon \sin{\nu t}) \equiv R_0/\alpha (t)$.
Transforming to a fixed spatial domain via
$\vec{y}\equiv \alpha(t)\vec{r}$, $y \equiv |\vec{y}| < R_0$, and 
renormalizing
the wavefunction $\phi ({\vec y}, t) \equiv \alpha^{-3/2}(t) 
\psi ({\vec r}, t)$ in order to preserve unitarity, we have
\be
i\hbar{\partial \phi \over \partial t}=H_0 \phi+ H_1(t) \phi \ \ ,
\la{e.scph}
\ee
where 
\be
H_1 (t) \equiv
\left( \alpha ^2(t) -1 \right)H_0 
-{\dot{R}(t)\over R(t)}\left(\vec{y}\cdot\vec{p}-i{3\over 2}\hbar \right) 
\ \ 
\la{e.scph1}
\ee
can be considered a small time-dependent perturbation if
$\epsilon$ and $\nu$ are small enough.

Since $H_1(t)$ commutes with $L^2$ and $\vec{L}$,  we 
can look for solutions that are eigenstates of the angular momentum.
This allows us to separate the angular dependence from the radial one in 
Eq.~\ref{e.scph} to obtain:

\be
\begin{array}{ll}
{\partial \over \partial t}\phi (y)&=i{\hbar\over 2m}\alpha ^2 (t)
\left[{\partial^2\over \partial y^2}+{2\over y}{\partial\over \partial 
y}-{l(l+1)\over y^2}\right]\phi(y) \\
&+ {\dot{R}(t)\over R(t)}
\left( y{\partial \over \partial y}+{3\over 2}\right)\phi(y) \ \ .
\end{array}
\la{e.scphr}
\ee

Using first-order perturbation theory, one can easily calculate
the coefficients of the 
solution's expansion in  terms of the unperturbed eigenstates.
If the initial state is chosen to be
$|i>=|n\!=\!k,l\!=\!0>$ ($\phi_{n,0}=\sqrt{2} n\pi j_0(n\pi y)$), we have
\be
\begin{array}{l}
c^0_n(t)=\delta_{n k}\\
c^1_n(t)={i\over \hbar}\delta_{n k} E_k\int^t_{0} 
dt'\left( 1- \alpha ^2 (t') \right) - \\
(-1)^{n-k}{2 n k\over n^2-k^2}(1-\delta_{n k})\int^t_{0} 
dt'e^{{i\over \hbar}(E_n-E_k)t'}{\dot{R}(t')\over R(t')} \ \ .
\end{array}
\la{eq.pert2}
\ee

The term due to $i\hbar{\dot{R}(t)\over R(t)}{3\over 2}$ is exactly
canceled out by the diagonal contribution of $-{\dot{R}(t)\over 
R(t)}\vec{y}\cdot\vec{p}$.
The last integral is analytically solvable for 
$\nu=\omega_{n k}=(E_n-E_k)/ \hbar$ and yields

\be
\begin{array}{l}
\int^t_0 dt'e^{i\omega_{nk}t'}{\dot{R}(t')\over R(t')}=
{\omega_{n k}t \over \epsilon}+ \cos\omega_{n k}t-1\\
-2{\sqrt{1-\epsilon^2}\over \epsilon}\left[\arctan\left({\epsilon+\tan 
\left({\omega_{n k}t\over 2}\right) \over \sqrt{1-\epsilon^2}}\right)
-\arctan\left({\epsilon\over \sqrt{1-\epsilon^2}}\right)\right] \\
+ i\left[\sin\omega_{n k}t-{1\over \epsilon}\ln(1+\epsilon\sin\omega_{n 
k}t)\right] \ \ . \end{array}
\la{eq.pert3}
\ee

The secular term $\omega_{n k}t/\epsilon$ in Eq.~\ref{eq.pert3} is a 
typical sign of a resonance. Notice that 
the secular term does not multiply a periodic function and the 
amplitude $\epsilon$ that we suppose to be small is at the 
denominator. We can easily check that this is not a problem if we 
make a Taylor expansion of 
$\arctan\left[ \left(\epsilon+\tan\left({\omega_{n k}t/ 2}\right)\right) 
/ \sqrt{1-\epsilon^2} \right]$ 
in powers of $\epsilon$ near $\epsilon=0$, 
since the zeroth-order term exactly cancels the secular term. 
However the increase of $c^1_n (t)$ in time remains.

\begin{figure}
\vspace*{0.15in}
\psfig{file=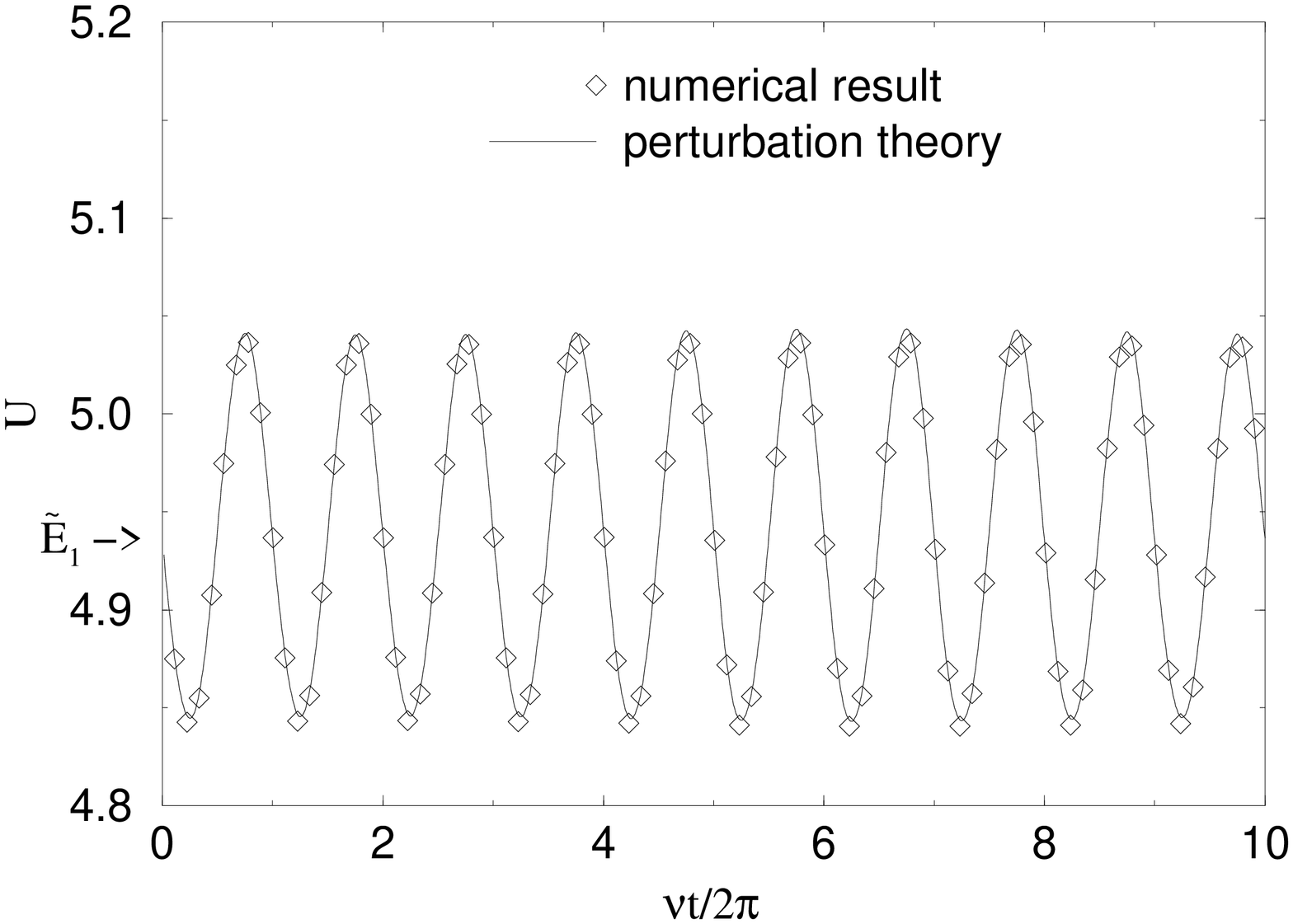,angle=270,width=9 cm}
\vspace*{0.2in}

\caption{Expectation value of the energy $U$ vs.~time (scaled by the 
oscillation period) for  
$\tilde{\nu}=7$ and $\epsilon=0.01$.  The initial state is chosen to
be $|n=1, l=0>$.}
\label{fig.etpr7l001}
\end{figure}

We can now calculate the expectation value of any observable as 
a function of time. 
We define the following dimensionless quantities:
\be
\begin{array}{l}
\tilde{E} \equiv m R_0^2E/ \hbar^2 \ , \\
\tilde{\nu} \equiv m R_0^2 \nu / \hbar \ .
\end{array}
\la{def.adim}
\ee
The perturbative results are in excellent agreement with the 
numerical ones when the cavity is oscillating out of the resonances. 
For example, at $ \tilde{\nu}\!=\!7 , \ \epsilon = 0.01$ 
the fluctuations of the energy (Fig.~\ref{fig.etpr7l001}) 
correspond almost exactly to those of ${1/ R^2(t)}$, as one can expect 
from a quasistatic approximation, even though our system is not quasistatic.
Even at high frequencies such as at $ \tilde{\nu}\!=\!90, \ 
\epsilon=0.01$, the first-order
perturbative results are still acceptable (Fig.~\ref{fig.hifreq}a). 
Notice that in this case the energy is 
shifted up slightly and its fluctuations in time are smaller. 
This is due to the fact that the system is no longer able to follow the fast 
oscillations of the walls, and consequently the fluctuations as well as the
value of the r.m.s.~radius $R_s \equiv <(y/R_o)^2>^{1/2}$
are  suppressed slightly (see Fig.~\ref{fig.hifreq}b).

At resonances, the perturbative approach 
breaks down and gives only an indication that a resonance exists.
In order to study these resonances we solved 
the Schr\"{o}dinger equation numerically, using a
unitary numerical algorithm \cite{goldb}. 
For $\tilde{\nu}=\tilde{E}_2-\tilde{E}_1$, we calculated the 
expectational values of the energy $U \equiv <\tilde{E}>$ and $R_s$,
choosing $|n=1,l>$ as the initial state.
In Fig.~\ref{fig.averes} 
we plotted the results for $l\!=\!0$ and $l\!=\!1$ ($l\!=\!0 \, , 
\, \tilde{E}_2-\tilde{E}_1\!=\!14.8044 \, ; \, l\!=\!1 \, , \, 
\tilde{E}_2-\tilde{E}_1\!=\!19.7444$) 
and two different values of 
$\epsilon$. The values for $\tilde{\nu}=7$ are also plotted for comparison.
The drastic change of behaviour of the system at the resonant 
frequency is evident even for very small amplitudes such as $\epsilon=0.001$.

At resonances the maximum expectation value of the energy, 
$U_{\rm max}$, varies as a function of $\epsilon$ 
because of the trivial adiabatic factor $\alpha^2(t)$ and, more 
importantly,  non-trivial excitation processes.
In  Fig.~\ref{fig.peaksa} we show ${\rm max}[\alpha^{-2}(t)U]$ 
vs.~$\epsilon$. 
For very small $\epsilon$ ($\epsilon < 0.002$),
the perturbation is not strong enough and the probability
of exciting the second eigenstate never reaches $1$. 
The expectation value of the energy saturates (and equals $\tilde{E}_2$) 
for $0.006<\epsilon<0.1$.  In this regime, 
the frequency dependence of the  energy maxima is well fitted by 
a Breit--Wigner function: \mbox{$U_{\rm max}=
\tilde{E}_1+C/[ (\tilde{\nu}-\tilde{\nu}_0)^2 +\Gamma^2/4]$},
and the width $\Gamma$ increases linearly with $\epsilon$ up
to $\epsilon \approx 0.1$.
For $\epsilon>0.1$, even higher states are excited.

Projecting the numerical solution on the eigenstates of the static system
we found the expected result that for $\epsilon < 0.1$, 
the resonant dynamics is dominated by the lowest two eigenfunctions.
This fact allows us to study the resonating system as a two-level system. 
In this case the differential 
equations for the coefficients reduce to:
\be
\dot{c}_1=-{i\over 
\hbar}\left[V_{11}(t)c_1+V_{12}(t)e^{-i\omega_{21}t}c_2\right] \ , \la{eq.21ev0}
\ee
\be
\dot{c}_2=-{i\over \hbar}\left[V_{21}(t)e^{i\omega_{21}t}c_1+V_{22}(t)c_2\right] 
\; , \la{eq.2lev1}
\ee
where $V_{ij}(t)\equiv <i|H_1(t)|j>$. 
Using the fact that $c_i(t)$ changes little in a period 
$T={2\pi}/\omega_{21}$,  we can average Eq.~\ref{eq.21ev0} and
Eq.~\ref{eq.2lev1} over a 
period to cast them into two coupled first-order ODE's with constant 
coefficients \cite{dodonov}:
\be
\dot{c}_i = \sum_j W_{ij} c_j \ \ .
\ee
Neglecting higher order terms in $\epsilon$, we have 
\be
W_{11}=W_{22}=0  \ , 
\ee
\be
W_{21}=-W_{12}= {4\left(1-\sqrt{1-\epsilon^2}\right)\over 3\epsilon} 
\omega_{21} \equiv  \Omega \ \ .
\la{eq.2lev2}
\ee
The system can then be diagonalized easily, giving
$c_1(t)=\cos\Omega t $ and $c_2(t)=\sin\Omega t $.

When $\epsilon \ll 1$ then $\Omega \simeq 2 \omega_{21} \epsilon/3$ and 
the period 
of the resonance $\lim_{\epsilon \rightarrow 0} T_r = {{2\pi} /\Omega} = 
\infty$. 
In the other limit when $\epsilon \rightarrow 1$ then $\Omega \rightarrow 
{4/3} \: \omega_{21}$, but in this case our assumption that 
$c_i(t)$ changes little in a period is no 
longer true and the averaging method no more valid.
In Fig.~\ref{fig.2lev1} we plot the expectation value of the energy 
$U=\alpha^2(t)(\tilde{E}_1 \cos^2 \Omega 
t +\tilde{E}_2 \sin^2 \Omega t)$ and compare it with the numerical results. 
For amplitudes $0.005 < \epsilon < 0.1$ the agreement is excellent. 

The matrix $W_{ij}$ can be written
as $-i\Omega \sigma_2$, where $\sigma_2$ is the second Pauli matrix.
It follows that the vector formed by the coefficients $c_1$ and $c_2$ 
behaves like the spinor of a spin $1/2$ particle in a magnetic field along the 
$\hat{\jmath}$ axis:
\be
i{\partial |\Psi> \over \partial t} = \Omega \sigma _2 |\Psi> \ \ .
\la{eq.spinan2}
\ee
Therefore, if the initial state of the particle inside the oscillating 
cavity is one of the two eigenstates involved in the resonance, 
which corresponds to an eigenstate of $S_z$, the evolution of the system
will be a precession of $<\vec{S}>$ around the $\hat{\jmath}$ axis. 
On the other hand, if the initial state corresponds to an 
eigenstate of $S_y$ we will obtain 
a stationary solution: $|\Psi(t)> = e^{\mp i\Omega t}|\Psi(0)>$,
which translates to
\be
\phi_{\pm}(y,t)=\sqrt{\alpha^3 (t)\over 2} 
e^{\mp i \Omega t}  
\left[e^{-i {E_1\over \hbar} t} \phi_1(y) 
\pm i e^{-i {E_2\over \hbar} t} \phi_2(y) \right] .
\la{eq.psiper1}
\ee
The wavefunction in Eq.~\ref{eq.psiper1} is periodical with 
period $T=2\pi/\omega_{21}$:
\be
\begin{array}{lll}
\phi_{\pm}\left(y,t+T\right)
=e^{i\theta} \phi_{\pm}\left(y,t\right) \ \ ,
\end{array}
\la{eq.psiper3}
\ee
where $\theta \equiv -2\pi[E_2/(\hbar\omega_{21}) \pm 
4(1-\sqrt{1-\epsilon^2})/3\epsilon]$.

We calculated numerically the solution choosing as initial function
one of the two of Eq.~\ref{eq.psiper1} at $t=0$, and in 
Fig.~\ref{fig.2lev1} 
we show the resulting $U$.  Although 
$\alpha^2(t) U(t)$ is not strictly constant 
its variation is considerably smaller compared to other solutions.
It is remarkable that such a highly dynamical system can show a
quasi-stationary behaviour.

For $\epsilon > 0.1$ the two-level approximation starts to break down. 
For $\epsilon = 0.15$ the third and fourth 
eigenstates become as important as the first two, and even more states
are involved as one increases $\epsilon$ further.
The behaviour of the system changes drastically for $\epsilon > 0.1$,
and we even observe the 
emergence of several new resonances that seem to have no straightforward
explanation in terms of the 
unperturbed eigenstates. In Fig.~\ref{fig.non} we show the maxima of 
$\alpha^2 (t) U(t)$ computed numerically for several driving frequencies
choosing as initial state $|n=1,l=0>$. The resonance at 
$\nu=\omega_{21}$ is indicated,
and it is much broader and smaller in amplitude compared to the new 
non-trivial resonances.
It is interesting to note that even at these new resonances,
the coefficients of the 
expansion in the static eigenstates are still approximately periodic.
It may be possible to understand these new resonances 
for $\epsilon > 1$ by including a few more levels in the two-level
approximation.
However the complexity of the system in this case warrants further study.

For $\epsilon < 0.005$
the two-level approximation fails again; it continues to give the maximum 
of the expected energy as $\tilde{E}_2$, typical of two-level systems, 
while in the complete system the  energy maximum decreases as 
$\epsilon$ is reduced. Also, the two-level approximation gives a period 
of the resonance $T_r$ greater than that of the complete system.

We emphasize that the resonances we studied here are 
caused exclusively by the motion of the cavity wall, since the system 
has no interaction with electromagnetic fields. 
Another interesting feature of our system is the independence of
its dynamics on $R_0$ except for the rescaling of the oscillating frequency.

It is also possible to consider a real system, 
hence with the electromagnetic interaction,
in which an ``oscillating-cavity'' resonance occurs but the Rabi 
resonances do not. In fact, to observe Rabi resonances we need a cavity with 
radius $R_0$ such that the fundamental frequency of the electromagnetic field 
$\nu_0={{2\pi}c/R_0}$ is equal to the difference between two energy levels,
$E_n-E_k \propto \hbar^2\pi^2/2mR_0^2$. It is hence 
not difficult to choose an $R_0$ such that the Rabi resonances are not 
excited. In practice though, maintaining a stable mechanical oscillation
with frequencies higher than some MHz is difficult.

For simplicity we have only considered a spherically symmetric cavity 
with perfect wall.  However, we conjecture that the resonances should not be 
too sensitive on the symmetry 
of the perturbation and on the detailed shape of the potential 
as long as the matrix element $V_{12}$ (see Eq.~\ref{eq.2lev1}) 
is different from zero.  One possibility is to use a microcrystal 
of conducting material with separations between the levels inside 
the conduction band of the order 
of $10^{-11}$ eV ($ \sim 100$ kHz). Forcing the crystal to vibrate 
at one of the resonant frequencies should excite many
of the Fermi level electrons, which decay by  emitting radiowaves.
A second way could be to use a system with several, almost equispaced, 
energy levels. At 
a resonant frequency the particle, an electron or a trapped atom for example,
absorbs energy from the driving oscillation to jump from one 
level to the next one and so on, as long as the resonance condition 
$\tilde{\nu}\simeq \tilde{E}_{n+1}-\tilde{E}_n$ is satisfied. 
In  this way the frequency of the emitted quanta
can be higher than the oscillation frequency, 
making them distinguishable from the electromagnetic noise due to 
dipole radiation at the driving frequency. 

In a further study
we will consider a system with many equispaced energy levels and 
analyze the increase in energy with time. Ideally from 
such a system one can get quanta of frequency much
higher than the driving frequency, and this is a major
difference compared to the cavity QED situation, where at
resonances typically a great increase in the number of photons with 
the {\it same} frequency as the driving force is expected.

We thank Dr.~C.~K.~Law for his suggestion of the two-level approximation.
This work is partially supported by the Hong Kong Research Grants Council 
grant CUHK 312/96P and a Chinese University Direct Grant 
(Project ID: 2060093).

\begin{figure}
\vspace*{0.3in}
\psfig{file=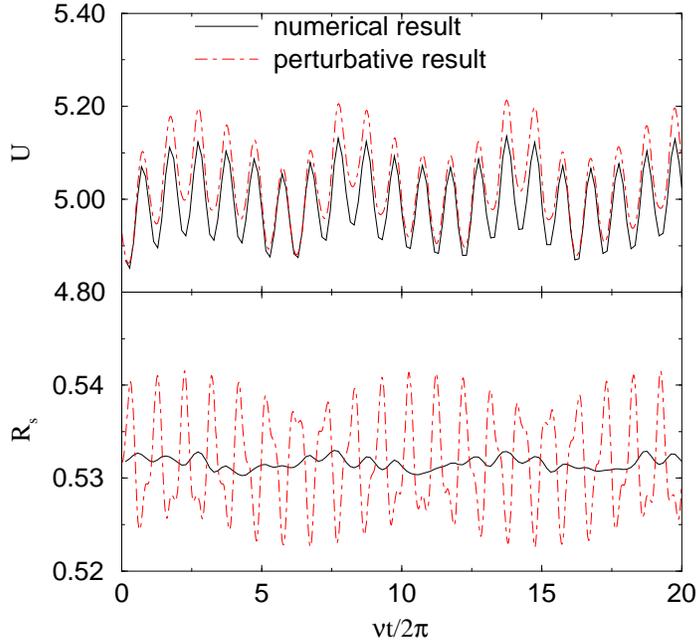,angle=0,width=9 cm}
\vspace*{0.2in}

\caption{(a) Comparison of energy calculated with the perturbative and 
numerical methods at a
high oscillation frequency ($\tilde{\nu}=90, \; \epsilon =0.01$); (b) 
r.m.s.~radius $<(y/R_0)^2>^{1/2}$ for high ($\tilde{\nu} = 90$, solid line)
and low ($\tilde{\nu} = 7$, dashed line) frequencies, both with $\epsilon
= 0.01$.}
\la{fig.hifreq} 
\end{figure}

\begin{figure}
\vspace*{0.25in}
\psfig{file=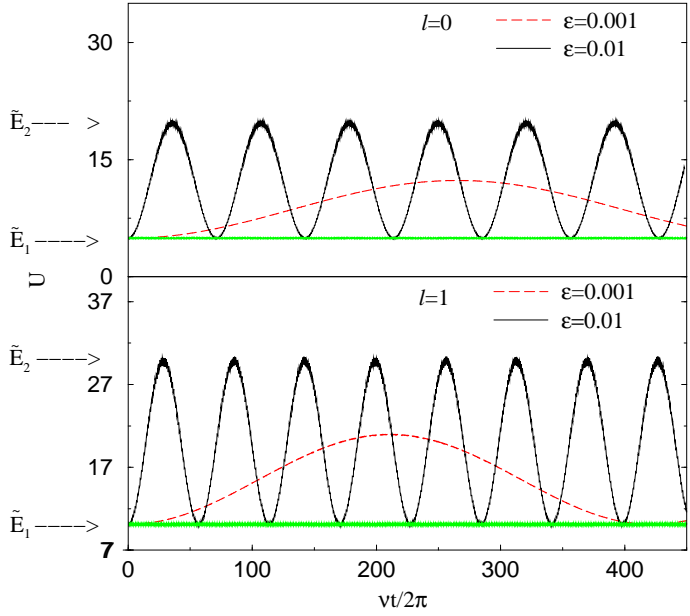,angle=270,width=9 cm}
\vspace*{0.2in}

\caption{Same as Fig.~1, but for 
(a) $l=0$, $\tilde{\nu}= 14.8044$ and (b) $l=1$, $\tilde{\nu} = 19.7444$
with $\epsilon = 0.001$ (dashed lines) and $\epsilon = 0.01$ (solid lines). 
The arrows indicate the two levels $\tilde{E}_1$ and $\tilde{E}_2$.
For comparison we also plotted the dependence for $\tilde{\nu}=7$, which show
up as flat lines near $\tilde{E}_1$. } 
\label{fig.averes}
\end{figure}

\begin{figure}
\vspace*{0.2in}
\psfig{file=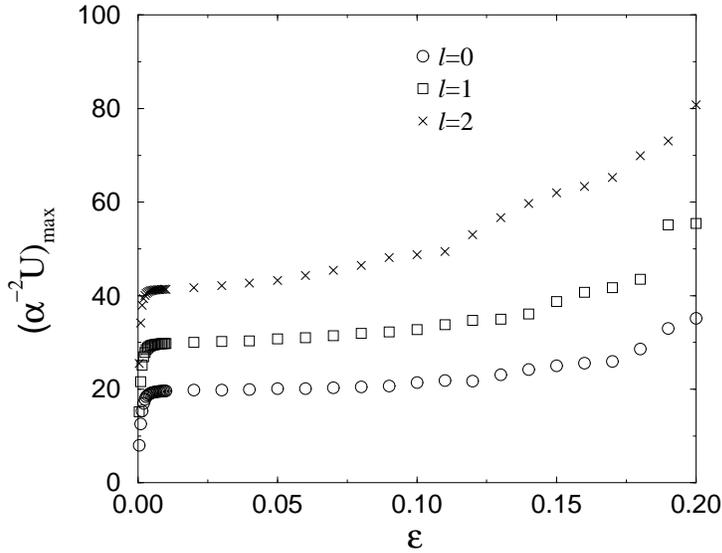,angle=0,width=9 cm}
\vspace*{0in}

\caption{Maximum expectation value of the energy $U_{\rm max}$
scaled by the ``trivial factor''
$\alpha ^2 (t) = (R_0/R(t))^2$ vs.~driving amplitude $\epsilon$
for $\tilde{\nu}=\tilde{E}_2-\tilde{E}_1$.}
\label{fig.peaksa}
\end{figure}

\begin{figure}
\vspace*{0.3in}
\psfig{file=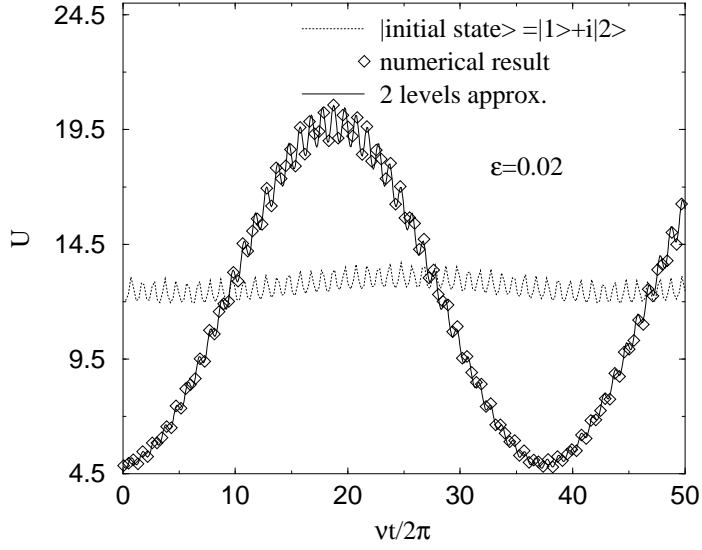,angle=0,width=9 cm}
\vspace*{0.2in}

\caption{Same as Fig.~1, but calculated with the
two-level approximation (solid line), for $l\! =\! 0$, 
$\epsilon \! =\! 0.02$, $\tilde{\nu}=\tilde{E}_2-\tilde{E}_1=14.8044$. 
The numerical results are also shown as diamonds. The dotted line shows
the numerical result for the special case when $|i>=1/\protect\sqrt{2} 
(|1>+i|2>)$} 
\label{fig.2lev1}
\end{figure}

\begin{figure}
\vspace*{0.3in}
\psfig{file=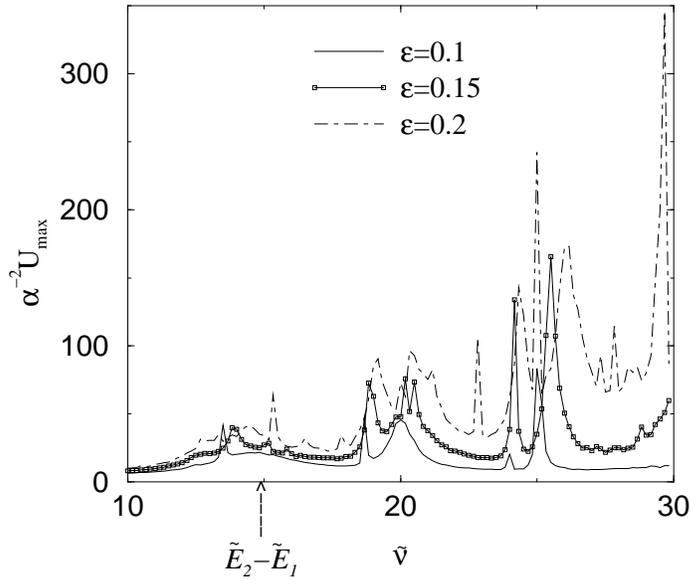,angle=0,width=9 cm}
\vspace*{0.2in}

\caption{Maximum expectation value of the energy 
$U_{\rm max}$ scaled by the ``trivial factor''
$\alpha^2(t)=(R_0/R(t))^2$ vs.~the driving frequency for $l=0$,  
$\epsilon = 0.1$ (solid line), 0.15 (squared line), and 0.2 (dashed 
line). Several nontrivial resonances appear.} \label{fig.non}
\end{figure}

\end{document}